\documentstyle[12pt]{article}
\makeatletter
\gdef\@pubnumber{\null}
\long\def\pubnumber#1{\gdef\@pubnumber{DAMTP 94-87}}
\def\@makepub{\vbox to \z@{\hbox to
\textwidth{\hfill\llap{\parbox[t]{0.33\textwidth}{\raggedleft\@pubnumber}}}%
\vss}}
\def\@maketitle{\newpage
\@makepub \null
\vskip 2em \begin{center}
{\LARGE \@title \par} \vskip 1.5em {\large \lineskip 0.5em
\begin{tabular}[t]{c}\@author
\end{tabular}\par}
\vskip 1em {\large \@date} \end{center}
\par
\vskip 1.5em}

\pubnumber{1}

\begin{document}

\title{Evolution of Fields in a
Second Order Phase Transition}
\author{Adrian Martin and Anne-Christine Davis\thanks{and King's
	College, Cambridge} \\
	\\
	{\normalsize DAMTP,}\\
	{\normalsize{Cambridge University,}}\\
	{\normalsize{Silver St.,}}\\
	{\normalsize{Cambridge,}}\\
	{\normalsize{CB3 9EW,}}\\
	{\normalsize{U.K.}}\\
	\\
	{\normalsize and}\\
	\\
	{\normalsize Isaac Newton Institute for Mathematical Sciences,}\\
	{\normalsize{Cambridge University,}}\\
	{\normalsize{Cambridge,}}\\
	{\normalsize{CB3 0EH,}}\\
	{\normalsize{U.K.}} }

\maketitle

\begin{abstract}
We analyse the evolution of scalar and gauge fields during a second
order phase transition using a Langevin equation approach. We show
that topological defects formed during the phase transition are stable
to thermal fluctuations. Our method allows the field evolution to be
followed throughout the phase transition,
for both expanding and
non-expanding Universes.
The results verify the Kibble mechanism
for defect formation during phase transitions.

\end{abstract}
\pagebreak

\section{Introduction}

Why there is so little anti-matter in the Universe, and why the
matter coalesced in the way it did are two of the major problems
facing cosmology. Predictably, both have attracted a great deal of
attention spawning a panoply of explanations and theories. Some of these
theories involve objects known as topological defects\cite{kibb}, regions of
trapped primordial vacuum, an example of which is the
cosmic string. A string approach is an
appealing one since it can be used to address both questions: the wake
left by  strings moving through the Universe can produce fluctuations
which may lead to the accretion of matter into large scale
structures\cite{bt}\cite{vish},
whilst their interaction with particles and the
decay of string loops can provide mechanisms
leading to baryon number violation\cite{bviol1} and the observed
matter bias\cite{bviol2}.
Hence, it is important to understand how these strings
form, both to predict how many we
can expect to have been created and how likely the above  processes
are.
The formation of topological defects is thought to proceed via
the Kibble mechanism\cite{kibb}.

Modern particle physics and the hot big bang model suggest that
as the Universe cooled it underwent several phase
transitions in which the symmetry of the vacuum was broken into a
successively smaller, and smaller group. During such a transition, it
is possible for fields to acquire non-zero vacuum expectation values.
How they do this depends on the order of the transition.

If we consider a transition where a U(1) symmetry is broken, then
following the transition all points in space will have a physically
identical, non-zero vacuum expectation value, the only variation being
in the difference in phase between any two points.

By causality, we expect the phases to be uncorrelated on
distances greater than the horizon length, and so there is a finite
probability that the phase along a closed path through a number of
Horizon volumes will wind through some multiple of $2\pi$ - an
indication that the loop contains a string. In practice two points do
not need to be separated by a horizon length for their phases to be
uncorrelated. This should also be true if they are a
thermal correlation length (defined later) apart;
usually a considerably
smaller distance than the horizon size.

This is the Kibble mechanism, and it relies upon the so-called geodesic
rule, which is that in passing between two domains of different phase,
the phase will follow the shortest route. In the global case this has
been verified both numerically\cite{ye} and
experimentally\cite{exp1}\cite{exp2},
though, for a  local symmetry it has been
argued\cite{rusri} that the presence of
gauge fields may influence the path the
phase takes, and may actually prevent it from following a `geodesic'.
More recently, however, work has been done suggesting that, despite
this, the
geodesic rule holds in the local case for a first order
phase transition\cite{hab}.

As the
temperature falls below the critical temperature, for a while it is
still possible for thermal fluctuations to restore the  broken
symmetry, and hence erase any topologically interesting configuration
present at the time. The
point at which it is thought that
this ceases to be possible is referred to as the
Ginzburg temperature, $T_{G}$, and is found by equating the free
energy with the thermal energy (for such a restoring fluctuation will
have a high probability while the former is considerably less
than the latter). Brandenberger and Davis\cite{rac} demonstrated that
given certain constraints on the parameters, the ratio of fluctuations
in the scalar field to the background is less than one beneath a
temperature just below the
Ginzburg temperature, regardless of whether gauge fields are, or are
not, present. This means that topologically non-trivial configurations
arising from thermal fluctuations will become stable to such
fluctuations just under the Ginzburg temperature. This adds weight to the
arguments in favour of the Kibble mechanism.

However, since Brandenberger and Davis considered a linearized model,
only valid for the short time immediately following a spinodal
decomposition, before non-linear effects start to
dominate, their analysis only holds for early times. To study the
evolution of the fields at later times it is necessary to include the
non-linearities. A flexible way to do this is to study the
Langevin equation associated with the classical field
equations\cite{bfmcg}.
This
is the purpose of this paper.

By studying the Langevin equation for the system,
we derive an equation for the probability distribution of the fields,
$P(\phi_{i},A_{i}^{\mu},t)$ which we use to analyse the evolution
of the expectation values of the classical fields coupled to a
thermal bath. This enables us to
study not only the stability of configurations to fluctuations at and below
the Ginzburg temperature, but also the long time evolution of the
fields. The flexibility of this method is demonstrated by the ease
with which it is modified to include the expansion of the Universe.

Our method seems to be the only one that allows the study of the effect that
thermal fluctuations have on the development of the field, and the
stability of defects formed, {\em throughout} the phase transition. Other
methods either concentrate on the start of the phase transition, or
near its completion.

\section{Global Symmetry}

Although our ultimate aim is to study the case of an expanding
universe with broken local symmetry, it is beneficial, for several reasons,
to start with the more
straightforward case of a global, non-expanding model. Firstly, since
the Kibble mechanism has already been verified for this case, we
know that any topologically non-trivial configurations present
should be stable to fluctuations below $T_{G}$,
and hence we have a benchmark to check
our results against. Although this provides a useful test of our
method, it is
by no means a proof of its validity. Secondly, since the amount of
algebra involved is very dependent on the number of fields present,
the global, non-expanding  case provides the simplest,
and hence clearest, demonstration
of the method used throughout.

Consider the $U(1)$ toy model
\begin{equation}
	{\cal{L}}=(D_{\mu}\phi)^{\dagger}(D^{\mu}\phi)
		-V({|\phi|}^2)
		-\frac{1}{4}F_{\mu\nu}F^{\mu\nu}
\end{equation}
where $\phi$ is a complex scalar field, $A_{\mu}$ is the U(1) gauge
connection (taken to be zero for now)
and $D_{\mu}=\partial_{\mu}-ieA_{\mu}$. We adopt an effective
potential of the form
\begin{equation}
	V({|\phi|}^2)=\frac{\lambda}{4}(|\phi|^{2}-\eta^{2})^{2}
			+\frac{\tilde{\lambda}}{2}T^{2}|\phi|^{2},
\end{equation}
where $\tilde{\lambda}=(4\lambda+6e^{2})/12$ and
the temperature dependence reflects the fluctuations on a scale
smaller than some correlation length, defined later.
For sufficiently high temperatures this is symmetric about a global
minimum at zero.
However, as the temperature passes through some critical temperature,
$T_{C}=( \lambda/\tilde{\lambda})^{\frac{1}{2}}\eta$,
the system undergoes a second order phase transition,
breaking the U(1) symmetry,
with new
minima appearing at
$|\phi|^{2}=\eta^{2}\left(1-T^{2}/T_{C}^{2}\right)$.
Any two points in the new vacuum will now
have non-zero vacuum expectation values of equal moduli but random phase.

Setting $\phi=\rho\exp{(i\alpha)}$,
our equations of motion for $\rho$ and $\alpha$ are
\begin{equation}
	\begin{array}{r}
	\partial_{\mu}\partial^{\mu}\rho-
	(\partial_{\mu}\alpha\partial^{\mu}\alpha) \rho+
	\frac{dV(\rho^{2})}{d\rho^{2}} \rho
	=0,\\
	\\
	\partial_{\mu}\partial^{\mu}\alpha+
	2\partial^{\mu}\alpha\partial^{\mu}\rho/\rho =0.\\
	\end{array}
	\label{octopus}
\end{equation}
We assume that $\alpha$ is time independent and varies spatially
over some length-scale $2\pi/k_{\alpha}$. Setting $R=\dot{\rho}$ we obtain
\begin{eqnarray}
	\dot{R}+F\rho  = 0 & , &
	\dot{\rho}-R  =  0,\\
	\label{2eq}
	\nonumber
\end{eqnarray}
where
\begin{equation}
	F=\left(k_{C}^{2}-
	\frac{\lambda}{2}\eta^{2}+
  	\frac{\tilde{\lambda}}{2}T^{2}-
	(k_{\alpha}\alpha)^{2}+
	\frac{\lambda}{2}\langle\rho^{2}\rangle\right),
\end{equation}
and $k_{C}$ is explained below.
Note that we have replaced $\lambda\rho^{3}/2$ with
$\lambda\langle\rho^{2}\rangle\rho/2$ c.f the mean square approximation
to make the resulting equations more accessible.

For the purpose of this analysis we consider an initial configuration
varying spatially on a scale of the correlation length. Since we do
not want to become embroiled in a discussion of effects due to
fluctuations on scales shorter than the thermal correlation length,
$\xi_{G}=1/[\eta\sqrt{\lambda(1-T_{G}^{2}/T_{C}^{2})}]$
, we consider a coarse-grained field where we have integrated out all modes
associated with such. This leads to the effective
potential mentioned earlier. Hence, if we were to perform a Fourier
decomposition, then it would be of the form
$\rho=\sum_{k\leq k_{c}}\rho_{k}\exp{(i\underline{k}.\underline{x})}$
for some $k_{c}\sim 1/\xi_{G}$. We also assume that the mode
corresponding to $k_{c}$ dominates, (which we later show to be
self-consistent) and investigate a configuration
with a length scale $2\pi/k_{C}$.
By (\ref{octopus}) we see that the earlier
assumption that $\dot{\alpha}=0$ requires
 $k_{\alpha}=2k_{c}$.

To incorporate thermal fluctuations into our model, we modify (\ref{octopus})
such that the equations describing
the evolution of $R$ and $\rho$ over some small time interval $\delta
t$ are
\begin{eqnarray}
	R(t+\delta t) & = & R(t)-\delta t F\rho(t) +\delta R,\\
	\rho(t+\delta t) & = & \rho(t) + \delta tR(t) +\delta\rho,\\
	\nonumber
\end{eqnarray}
where $\delta\rho$ and $\delta R$ are the thermal fluctuations in
$\rho$ and $R$ respectively.

Defining $\rho_{\delta t}=\rho(t+\delta t)$, $\rho=\rho(t)$, and
similarly $R_{\delta t}$ and $R$, we can write
\begin{eqnarray}
	P(\rho_{\delta t},R_{\delta t},t+\delta t) & = &
	\int d(\delta \rho) d(\delta R)
	P_{1}(\delta \rho)P_{2}(\delta R)
	\nonumber\\
	 & & \times P(\rho_{\delta t}-\delta tR_{\delta t}-\delta\rho,
	R_{\delta t} +\delta tF\rho_{\delta t}-\delta R,t)
	\nonumber\\
	 & & \times\frac{\partial}{\partial\rho_{\delta t}}
	(\rho_{\delta t}-\delta tR_{\delta t}-\delta\rho)
	\nonumber\\
	 & & \times\frac{\partial}{\partial R_{\delta t}}
	(R_{\delta t} +\delta tF\rho_{\delta t}-\delta R),\\
	\nonumber
\end{eqnarray}
where $P_{1}$ and $P_{2}$ are the probability measures for
$\delta\rho$ and $\delta R$ respectively.

Since $\delta\rho$ and $\delta R$ are random fluctuations, we may
assume that
$\langle\delta\rho\rangle$=$\langle\delta R\rangle=0$.
Expanding the integrands as Taylor series, we find, after
considerable algebra, that
\newline
\begin{eqnarray}
	P_{\delta t} & = & P
	-\delta t \frac{\partial}{\partial\rho}(RP)
	+\delta t \frac{\partial}{\partial R}(F\rho P)\nonumber\\
	 & & +\frac{\partial}{\partial\rho}
	\left(
	\frac{1}{2}\langle\delta\rho^{2}\rangle
	\frac{\partial P}{\partial\rho}
	\right)
	+\frac{\partial}{\partial R}
	\left(
	\frac{1}{2}\langle\delta R^{2}\rangle
	\frac{\partial P}{\partial R}
	\right)\\
	\nonumber
\end{eqnarray}
from which, assuming that $\delta\rho$ and $\delta R$ are independent of
$\rho$ and $R$, we obtain the
following differential equation for P:
\begin{eqnarray}
	\frac{\partial P}{\partial t} & = &
	-\frac{\partial}{\partial\rho}(RP)
	+\frac{\partial}{\partial R}(F\rho P)\nonumber\\
	 & &
	+\frac{1}{2}
	\frac{\langle\delta\rho^{2}\rangle}{\delta t}
	\frac{\partial P}{\partial\rho}
	+\frac{1}{2}\frac{\langle\delta R^{2}\rangle}{\delta t}
	\frac{\partial P}{\partial R}.\\
	\nonumber
\end{eqnarray}
One interpretation of this equation is as follows. If we move the
first two terms on the right hand side over to the left, then we have
a full derivative of $P$. Liouville's theorem states
that for a closed system this derivative should be zero. However, our
system is coupled to a thermal bath and so there is a flow of
probability between the two, as demonstrated by the two non-zero noise
terms, due to the bath, on the right hand side.

This equation is clearly a rather forbidding equation to solve analytically.
However, we can use it to derive equations governing the quadratic
moments
\begin{eqnarray}
	\langle\rho^{2}\rangle =
	\int d\rho dR P(\rho,R,t)\rho^{2}, &
	\langle\rho R\rangle =
	\int d\rho dR P(\rho,R,t)\rho R, &
	\langle R^{2}\rangle  =
	\int d\rho dR P(\rho,R,t)R^{2}.\nonumber\\
	\nonumber
\end{eqnarray}
We choose to investigate these moments because their equations form a
closed system with the mean field approximation we have taken.

As a modification we set
\begin{equation}
	\begin{array}{ccccccc}
	u=\langle\rho^{2}\rangle/\eta^{2} & , &
	v=\langle\rho R\rangle/\eta^{3} & , &
	w=\langle R^{2}\rangle/\eta^{4} & , &
	\tau=\eta t,\\
	\end{array}
\end{equation}
since this normalises $u$, $v$ and $w$, and gives the equations for
the moments in
a form where the relative sizes of terms are much more apparent;
\begin{eqnarray}
	\dot{u} & = & 2v+\delta_{1},\label{neq1}\\
	\dot{v} & = & - f_{o}u - \frac{1}{2}\lambda u^{2} + w^{2},
		\label{neq2}\\
	\dot{w} & = & - 2f_{o}v - \lambda uv +
		\delta_{2},
		\label{neq3}\\
	\nonumber
\end{eqnarray}
where $^{.}$ denotes $\frac{d}{d\tau}$,
\begin{eqnarray}
	\delta_{1}=
	\frac{1}{\eta^{3}}
	\frac{\langle\rho^{2}\rangle}{\delta t}, & &
	\delta_{2}=\frac{1}{\eta^{5}}
	\frac{\langle R^{2}\rangle}{\delta t}\\
	\nonumber
\end{eqnarray}
are the fluctuations and
\begin{eqnarray}
	f_{o} & = & \left(\frac{k^{2}}{\eta^{2}}
			-\frac{\lambda}{2}
			+\frac{\tilde{\lambda}}{2}\frac{T^{2}}{\eta^{2}}
			-\frac{(k_{\alpha}\alpha)^{2}}{\eta^{2}}\right).\\
	\nonumber
\end{eqnarray}

It should be noted that these equations are even nastier than
they look at first glance,
since $f_{o}$ contains a term proportional to $T^{2}$. However,
by assuming that the temperature varies at a much slower rate than the
fields (something which we will see is self-consistent later on) it is
possible to make some progress analytically.

For the time being, we consider the case when fluctuations are absent,
and treat $f_{o}$ as constant over a small time period.
After a bit of
substitution, we integrate the equations to obtain
\begin{eqnarray}
	\dot{u}^{2} & = & -\lambda u^{3}
		-4f_{o}u^{2}
		+\Lambda_{1}u
		+\Lambda_{2},\\
	\nonumber
\end{eqnarray}
where $\Lambda_{1}=4(w_{i}+f_{o}u_{i})+\lambda u^{2}_{i}$,
$\Lambda_{2}=4(v_{i}^{2}-u_{i}w_{i})$ and $u_{i}$, $v_{i}$ and $w_{i}$
are the initial values of $u$, $v$ and $w$.

Let the three roots of
the polynomial on the right hand side be
$\mu_{1}>\mu_{2}>\mu_{3}$. Hence
\begin{equation}
	\begin{array}{ccc}
	\mu_{1}\mu_{2}\mu_{3}=\frac{4\Lambda_{2}}{\lambda}, &
	\mu_{1}\mu_{2}+
	\mu_{2}\mu_{3}+
	\mu_{3}\mu_{1}=-\frac{4\Lambda_{1}}{\lambda}, &
	\mu_{1}+\mu_{2}+\mu_{3}=-\frac{4f_{o}}{\lambda}.\\
	\end{array}
\end{equation}
Taking initial conditions such that, at $T_{G}$,
$u_{i}=1-t_{C}/t_{G}=\lambda^{2}/[g(1+\lambda^{2}/g)]$,
which is the minimum of the effective potential
at that time, and $v_{i}=w_{i}=0$, we find  $\Lambda_{2}=0$,
implying that one of the roots is zero. Furthermore, $\Lambda_{1}$ and
$f_{o}$ are
found to be negative, so the two non-zero roots must be positive. Since
$\dot{u}^{2}$ is seen to be positive between $\mu_{1}$ and $\mu_{2}$
we expect $u$ to oscillate between these two.

We can also make an
estimate of the time period of these oscillations, $t_{P}$, since
\begin{equation}
	 t_{P}=
	\frac{4}{\sqrt{\lambda\mu_{1}}}K(\kappa),
\end{equation}
where
\begin{eqnarray}
	K(\kappa) & = &
	\frac{\sqrt{\mu_{1}}}{2}
	\int^{\mu_{1}}_{\mu_{2}}
	\frac{du}{\sqrt{(\mu_{1}-u)(u-\mu_{2})(u-\mu_{3})}},\nonumber\\
	\nonumber
\end{eqnarray}
is the first complete elliptic integral and
$\kappa=\sqrt{[(\mu_{1}-\mu_{2})/\mu_{1}]}$.
These results closely agree with the corresponding numerical calculations.
However, they are only valid for very small fluctuations, and so we
turn to a numerical approach for a more detailed analysis.

In order to study the effect of fluctuations on the evolution of the
fields, it is necessary to
make some estimate of the
fluctuation terms. To do this we imagine $\phi$ coupled
to some other field, $\psi$, in
thermal equilibrium, via an extra term in the Lagrangian of the form
\begin{eqnarray}
	{\cal{L}}_{I} & = & \frac{1}{2}g|\phi|^{2}|\psi|^{2}.\\
	\nonumber
\end{eqnarray}
By comparing the resulting equations of motion with those already
obtained we find that
\begin{eqnarray}
	\delta\rho=0 & , & \delta R=g\rho\psi^{2}\delta t.\nonumber\\
	\nonumber
\end{eqnarray}

Since $\psi$ is in thermal equilibrium, we also have that $\psi\sim
T$, with corresponding number density
$n_{\psi}  = 1.202g_{*}T^{3}/\pi^{2}$
where $g_{*}$ is the number of internal degrees of freedom
(107 for a Grand Unified Theory). Taking
$\delta t$ to be a typical interaction time, such that $\delta
t\sim n_{\psi}^{-1}\sigma_{I}^{-1}$, where $\sigma_{I}=\sim gk^{-2}$ is
the interaction cross-section, we find the following approximation for
the fluctuation terms
\begin{eqnarray}
	\delta_{1}=0 & , &
	\delta_{2}=
	g \left(\frac{k^{2}}{\eta^{2}}\right)\frac{T}{\eta}u.\\
	\nonumber
\end{eqnarray}
As expected, the size of the fluctuations decreases with temperature.

Before we can carry out a calculation,
we need to address the problem of choice of parameter values.
Since the model we are
considering is a global one, we see from the definition of the
potential that we must have  $\tilde{\lambda}=\lambda/3$.
We now demonstrate why we are justified in assuming that the mode
corresponding to $k_{C}$ dominates.
Figure 1 shows the evolution of the quadratic moment $u$,
(corresponding to $\langle \rho^{2}\rangle$) from the Ginzburg time
onwards, where we have taken $\lambda=0.1$,
fluctuation coupling, $g=\lambda/3$ and a range of different
wavelengths.
The most obvious feature is that
the mode varying with wavenumber $k_{C}$ dominates those with longer
wavelengths, consistent with our earlier assumption. However, one may
be slightly alarmed at the fact that at least two of the curves look
like they have no intention of converging to one (corresponding to
$\langle\rho^{2}\rangle\rightarrow\eta^{2}$) as one might expect. The
reasons for this are twofold, and both somewhat of our own creation.
The first is that in assuming that $\rho$ and $\alpha$ vary spatially
with fixed wavenumber, we alter the value of $\rho$ for which
$\dot{\rho}=0$ since we have essentially added two terms onto the
derivative of the potential. This has the effect of raising the
equilibrium value of $\rho$. The second is that we have to make an
arbitrary choice of $\alpha$ (taken throughout as $\alpha=1$),
which effectively scales $k_{\alpha}$,
and so has a similar effect to the first. Conversely it could also be
used to tune the expected equilibrium value to one by choosing a
sufficiently small value of $\alpha$.

This may raise questions over
the validity of this method for studying the evolution of fields.
However, the evolution of the fields is not qualitatively changed by
taking different values of $k_{C}$, $k_{\alpha}$ or $\alpha$ and so we
argue that, as an approximation, our approach is still of interest.

The potential we are using is, unfortunately, only a one loop
approximation, and hence is not
valid above
the Ginzburg temperature where higher loops dominate.
Our simulation therefore must run from the Ginzburg time
onwards, and so we can only investigate the stability of string
configurations to thermal fluctuations and not their formation.
We also only consider the case of a GUT phase transition, since we
expect one at a lower temperature, such as the Electro-weak phase
transition for instance, to be qualitatively the same, but slower due to
the larger value of $t_{C}$.
We take the coupling $\lambda$ to be between 0.01 and 1, and
$g\leq\lambda$.

Figure 2 shows the effect of the two couplings, $g$
and $\lambda$. The former controls the size of fluctuations, and it is
seen, in 2a and 2b,
that the larger $g$, the quicker the rise of the lower bound, so fluctuations
are, bizarrely, actually seen to help
stabilise configurations, on average, by damping oscillations in the
field.
They also cause the upper bound
to rise at a faster rate though this is not as pronounced, nor as
important to the stability of domain structures.

Fig.2c reveals that decreasing $\lambda$ decreases the frequency of
oscillations, and also the asymptotic value for $u$. The latter is
because $k_{C}\propto 1/\xi_{G}\propto\sqrt{\lambda}$ and we have
already noted that the value of $k_{C}$ effects the limiting value.
Finally, Fig.2d demonstrates that the effect of fluctuations decreases
dramatically with $\lambda$.

We see then that, since all curves move away from zero, any
topologically non-trivial configuration is stable from the Ginzburg
temperature onwards, though the fields may take a long time to
reach their equilibrium values. We also note that, in all cases
considered, the
oscillations occur on a much smaller timescale than the evolution
towards the equilibrium value; consistent with our earlier assumption.

\section{The Effect of Gauge Fields}

That the configurations formed in the above transition are stable
against thermal fluctuations is nothing new.
The Kibble mechanism for the global
case is already well accepted, since one can argue in favour of the
geodesic rule just by demanding that the path followed minimizes the
energy density. However, the presence of gauge fields may undermine
this, since their presence in the gradient energy,
$(D_{\mu}\phi)(D^{\mu}\phi)^{\dagger}$,
may make it equally favourable, energetically, to
follow a longer path.

Luckily, the method used for studying the global case works equally
well in the local one, the only drawback being a significant increase in the
amount of algebra that has to be done. We start by writing the
equations of motion in the form
\begin{eqnarray}
	\partial_{\mu}\partial^{\mu}\rho
	-e^{2}(q_{\mu}q^{\mu})\rho+
	\frac{\partial V(\rho^{2})}{\partial\rho^{2}}\rho
	 & = & 0,\nonumber\\
	\partial_{\mu}\partial^{\mu}q^{\nu}
	+2e^{2}\rho^{2}q^{\nu}
	+\frac{1}{e}\partial_{\mu}\partial^{\mu}\partial^{\nu}\alpha
 	 & = & 0,\nonumber\\
	\nonumber
\end{eqnarray}
where $q^{\nu}=A^{\nu}-\frac{1}{e}\partial^{\nu}\alpha$ (Note that
this is gauge invariant). Setting
$Q^{\mu}=\dot{q}^{\mu}$,
$R=\dot{\rho}$ and
$\Delta^{\nu}=\frac{1}{e}\partial_{\mu}\partial^{\mu}\partial^{\nu}\alpha$
for convenience, we find
\begin{equation}
	\begin{array}{lcrcrcrcr}
	R(t+\delta t) & = &
	R(t) & - &
	\delta tF_{2}(t)\rho(t) & + &
	\delta R, & &\\
	\rho(t+\delta t) & = &
	\rho(t) & + &
	\delta tR(t) & + &
	\delta\rho, & &\\
	Q^{\mu}(t+\delta t) & = &
	Q^{\mu}(t) & - &
	\delta tG(t)q^{\mu}(t) & + &
	\delta Q^{\mu} & - & \delta t\Delta^{\mu},\\
	q^{\mu}(t+\delta t) & = &
	q^{\mu}(t) & + &
	\delta t Q^{\mu}(t) & + &
	\delta q^{\mu}, & &
	\end{array}
\end{equation}
where
\begin{eqnarray}
	F_{2} & = &
	\left(
	k^{2}
	-\frac{\lambda}{2}\eta^{2}
	+\frac{\tilde{\lambda}}{2}T^{2}
	+\frac{\lambda}{2}\langle\rho^{2}\rangle
	-e^{2}\langle q^{2}\rangle
	\right),\nonumber\\
	G &  = &
	\left(k^{2}
	+2e^{2}\langle\rho^{2}\rangle\right),\nonumber\\
	\nonumber
\end{eqnarray}
and $\delta R$, $\delta\rho$, $\delta Q^{\mu}$, $\delta q^{\mu}$ are
the thermal fluctuations.

Proceeding in exactly the same manner as before,
we arrive, after
some very unpleasant algebra, at the equation for
$P(\rho,R,q^{\mu},Q^{\mu},t)$,
\begin{eqnarray}
	\frac{\partial P}{\partial t} & = &
	-\frac{\partial}{\partial\rho}(RP)
	+\frac{\partial}{\partial R}(F_{2}\rho P)
	-\frac{\partial}{\partial q_{\mu}}(Q_{\mu}P)
	+\frac{\partial}{\partial Q_{\mu}}(Gq_{\mu}P)
	+\Delta_{\mu}\frac{\partial P}{\partial Q_{\mu}}
	\nonumber\\
	  & &
	+\frac{1}{2}
	\frac{\langle\delta\rho^{2}\rangle}{\delta t}
	\frac{\partial^{2}P}{\partial\rho^{2}}
	+\frac{1}{2}
	\frac{\langle\delta R^{2}\rangle}{\delta t}
	\frac{\partial^{2}P}{\partial R^{2}}
	\nonumber\\
	 & &
	+\frac{1}{2}
	\frac{\langle\delta q_{\mu}\delta q^{\mu}\rangle}{\delta t}
	\frac{\partial^{2}P}{\partial q_{\mu}\partial q^{\mu}}
	+\frac{1}{2}
	\frac{\langle\delta Q_{\mu}\delta Q^{\mu}\rangle}{\delta t}
	\frac{\partial^{2}P}{\partial Q_{\mu}\partial Q^{\mu}}.
	\nonumber\\
	\nonumber
\end{eqnarray}
It is important to note that the sum over indices in the last two
terms is over all four indices, not the usual two.
Once more we see the violation of Liouville's theorem via the coupling
to the heat bath.

{}From this it is straightforward to obtain the equations governing the
quadratic moments.
Following the global method and defining $\tau$,
$u$, $v$, $w$ as before, plus
\begin{equation}
	\begin{array}{ccccc}
	x=\langle q^{2}\rangle/\eta^{2} & , &
	y=\langle q^{\mu}Q_{\mu}\rangle/\eta^{3} & , &
	z=\langle Q^{2}\rangle/\eta^{4}\\
	\end{array}
\end{equation}
we arrive at
\begin{equation}
	\begin{array}{cclcccl}
	\dot{u} & = & 2v+\delta_{1}, &  &
	\dot{x} & = & 2y+\delta_{3}, \\
	\dot{v} & = &
	- f_{2}u - \frac{\lambda}{2}u^{2} + e^{2}xu + w, &  &
	\dot{y} & = &
	- \left(\frac{k^{2}}{\eta^{2}}\right)x
	- 2e^{2}ux + z,\\
	\dot{w} & = &
	- 2f_{2}v - \lambda uv + 2e^{2}xv
	+ gu\left(\frac{k^{2}}{\eta^{2}}\right)\frac{T}{\eta}+\delta_{2}, & &
	\dot{z} & = &
	- 2\left(\frac{k^{2}}{\eta^{2}}\right)y
	- 4e^{2}uy+\delta_{4},\\
	\end{array}
\end{equation}
where
\begin{eqnarray}
	f_{2} & = & \left(\frac{k^{2}}{\eta^{2}}
			-\frac{\lambda}{2}
			+\frac{\tilde{\lambda}}{2}\frac{T^{2}}{\eta^{2}}
			\right),\\
	\nonumber
\end{eqnarray}
and $\delta_{1}$, $\delta_{2}$, $\delta_{3}$ and $\delta_{4}$ are
the thermal fluctuation terms.
Note that the terms involving $\Delta_{\mu}$ have integrated to zero.
Clearly we are not going to get too far with an analytic approach this
time, so we restrict ourselves to a numerical analysis.

As before,
our first preparation  is to calculate the fluctuation terms. Since
the coupling of $\phi$ to $\psi$ is independent of gauge fields, the
two new
fluctuation terms, $\delta_{3}$ and $\delta_{4}$ must be zero. Hence,
the only non-zero fluctuation term is $\delta_{2}$, which is unchanged.

We consider once more an initial domain structure of length scale
$\xi_{G}$, and take $e^{2}=40\lambda/3$, since we expect the gauge
coupling to dominate. Figure 3 shows the effect of varying $\lambda$
and $g$, the results being very similar to those observed in the
global symmetry case. Figures 3a and 3b once again demonstrate the
effect of increasing the size of the fluctuations; an increase in the
rate of growth of the lower bound
and a damping of oscillations, whilst Fig.3c reveals how
decreasing the size of the self-coupling decreases the frequency of
oscillations, the effect of fluctuations decreasing in a similar
manner (Fig.3d).

The most important feature however is that, as in the global case,
any non-trivial
domain structure present at $t_{G}$ is seen to be stable against
fluctuations at greater times. This reinforces the work by
Brandenberger and Davis\cite{rac}, and, similarly, the arguments in
favour of the Kibble mechanism.

\section{Including the Expansion of the Universe}

Until now we have ignored the expansion of the Universe,
which we would expect to damp the amplitude of any oscillations present.
To make the analysis more realistic it is necessary to include this
expansion.

\subsection{Global Symmetry}

Taking first the case with a global symmetry, such a modification is
straightforward, and leads to an extra term in the equation of motion
for $\rho$ proportional to the Hubble parameter $H$,
\begin{equation}
	\partial_{\mu}\partial^{\mu}\rho-
	\partial_{\mu}\alpha\partial^{\mu}\alpha \rho+
	\frac{dV(\rho^{2})}{d\rho^{2}} \rho
	=-3H\frac{\partial\rho}{\partial t}:
\end{equation}
as expected, a damping term.
The only effect this has on the equations for the quadratic moments
$u$, $v$ and $w$, is to add $-3Hv$ to the right hand side of
equation (\ref{neq2}), and $-6Hw$ to that of
(\ref{neq3}), though $f_{o}$ is now written as
\begin{eqnarray}
	f_{o} & = & \left[
			\frac{k^{2}}{\eta^{2}}
			\left(\frac{a_{0}}{a}\right)^{2}
			-\frac{\lambda}{2}
			+\frac{\tilde{\lambda}}{2}\frac{T^{2}}{\eta^{2}}
			-\frac{(k_{\alpha}\alpha)^{2}}{\eta^{2}}
			\left(\frac{a_{0}}{a}\right)^{2}\right]\\
	\nonumber
\end{eqnarray}
where $a$ is the expansion parameter, and $a_{0}$ its value at
$T_{G}$\footnote{Since $a(t)$ is an unphysical quantity, we can
without loss of generality take $a_{0}=1$.}.

Figure 4 shows the results for a small selection of values
to illustrate the effects of varying the different parameters. Once
more we consider an initial domain structure of length scale
$\xi_{G}$.
All four diagrams are seen to display the rapid damping due to the expansion of
the Universe.

Figs. 4a and 4b demonstrate the effect of fluctuations. In 4b, where
the fluctuations are suppressed, the lower bound on $u$ rises much
more slowly than in the unsuppressed case, Fig.4a, which actually
overshoots its asymptotic value of one before reconverging.
Hence, it is seen that fluctuations
actually make it less likely that a configuration will be erased,
agreeing with our non-expanding simulations. The
upper bound varies very little between the two.

Fig. 4c shows the effects of reducing the self-coupling; a longer
period of oscillation, a less dramatic initial growth and a much gentler
approach toward its asymptotic value. Fig. 4d demonstrates how for small
values of $\lambda$ the fluctuations have very little effect on the
evolution of the fields.

In summary, topologically non-trivial configurations are stable to
thermal fluctuations.
We also note that due to the scale factor now present in the equations of
motion, the effect of $k_{c}$ and $k_{\alpha}$ is rapidly damped out
so that in all expanding cases considered, $u$ converges on one,
corresponding to $\langle \rho^{2} \rangle$ tending to $\eta^{2}$, the
long-time minimum of the effective potential.

\subsection{Local Symmetry}

Now including gauge fields once more,
the equation of motion for $A_{\mu}$ is, predictably, modified in a very
similar way to that for $\rho$ when we include expansion, the new
version being
\begin{eqnarray}
	\partial_{\mu}\partial^{\mu}A^{\nu}+
	2e^{2}\rho^{2}A^{\nu}-2e\rho^{2}\partial^{\nu}\alpha & = &
	-3H\partial_{0}A^{\nu}.\\
	\nonumber
\end{eqnarray}
In addition, the equations for the
quadratic moments $x$, $y$ and $z$ acquire near
identical terms to those already acquired by those for $u$, $v$ and
$w$, the only difference being an extra factor of two in the former case.

Once more, we illustrate four different values of the parameters, for a
configuration varying on a correlation length scale.
As in the non-expanding case, we take
$e^{2}=40\lambda/3$ throughout.

Figures 5a and 5b demonstrate the effect of fluctuations. In Fig.5a, where
the fluctuation coupling is comparable to the self-coupling, the lower
bound to fluctuations is seen to rise much quicker (as was the case
for a global symmetry) than that in Fig.5b where the fluctuations are
suppressed. So much so in fact that it overshoots its expected limit,
though long
time studies show that it gradually bends back and converges to one.

In Fig.5c we see the effect of decreasing the self-coupling; a less
dramatic growth of the lower bound, whilst Fig.5d reveals, once more,
that the effect
of fluctuations decreases dramatically with $\lambda$.
Much the same as in the previous three cases.

Comparing figures 4 and 5, what we notice is that the presence of
gauge fields heavily
damps the initial growth leading to a lower upper bound and
consequently smaller oscillations.

\section{Conclusions}

By studying the Langevin equations for the classical fields we have
verified that for a U(1) model with broken global symmetry (whether
expanding or not), string
configurations formed during a second order phase transition, are
stable to thermal fluctuations below the Ginzburg temperature. We have
also shown this to be true in the case of a system with a local
symmetry, reinforcing earlier work\cite{rac} on the subject, and
lending further support to the Kibble mechanism for the formation of
topological defects.

The same method has also been used to study how the fields
evolve at late times, with the
scalar field gradually tending to its equilibrium value;
a process accelerated by the damping produced by an expanding
universe.
Indeed, our method tracks the evolution of the field {\em throughout}
the phase transition. Other methods are only able to consider early or
late times.
In addition, we have seen that thermal fluctuations
actually accelerate
the early evolution of the field, and damp the amplitude of
oscillations in the field as it tends to its asymptotic value,
making it even less likely that a fluctuation will destroy a
configuration.

This work is still an approximation however, since we have had to
assume $\dot{\alpha}=0$, leading to an arbitrariness in the asymptotic
value of the field in the non-expanding models. However, this should
not affect the stability of configurations, and in the expanding case
this problem is smoothed out anyway as the model scales with time.

Another approximation we have made is in neglecting the dissipation term
necessary when a source of fluctuations is
present\cite{diss1}\cite{diss2}. Since any dissipation term would have a
damping effect, it would only increase the stability of a
non-trivial domain structure, and so including it, a further
avenue of research, should only strengthen our results.  This and the
quantum field theoretical treatment are in progress\cite{ray}

Finally, the flexibility of the Langevin equation
approach\cite{bfmcg}
may make this
method suitable for a number of other applications, such as the study
of the evolution of seed magnetic fields following the breaking of a
non-Abelian symmetry\cite{vachy}, and the stability of defects to fluctuations
in
condensed matter systems such as $^{4}$He\cite{zurek}. Work on these subjects
is
in progress\cite{new}.

We are indebted to Robert Brandenberger and Ray Rivers for discussions
and suggestions.
This work is supported in part by P.P.A.R.C. and E.P.S.R.C.

\end{document}